\documentclass[11pt]{article}
\usepackage[top=2cm,left=2cm,right=2cm]{geometry}
\usepackage[numbers,square]{natbib}

\usepackage[italian,english]{babel}
\usepackage{epsfig}
\usepackage{hyperref}
\usepackage{amssymb}
\usepackage{amsfonts}
\usepackage{epsf}
\usepackage{rotating}
\usepackage{graphicx}
\usepackage{amsmath}
\usepackage{fancyhdr}
\usepackage{lineno}
\usepackage{subfigure}
\usepackage{babel}
\usepackage{graphics}
\usepackage{geometry}
\usepackage{pstricks}
\usepackage{color}

\makeatletter
\def\cleardoublepage{\clearpage\if@twoside
\ifodd\c@page
\else\hbox{}\thispagestyle{empty}\newpage
\if@twocolumn\hbox{}\newpage\fi\fi\fi}
\makeatother

\let\a=\alpha   \let\b=\beta      \let\d=\delta
    \let\h=\eta     
        
                 \let\r=\rho
\let\s=\sigma        \let\f=\phi

\def\a{\alpha}
\def\b{\beta}

\def\r{\rho}

\def\ds#1{#1\kern-1ex\hbox{/}}
\def\dsh{h\kern-1.2ex /}

\newcommand{\bea}{\begin{eqnarray}}
\newcommand{\eea}{\end{eqnarray}}

\def\nn{\nonumber}

\def\beq{\begin{equation}}
\def\eeq{\end{equation}}

\def\ba{\begin{eqnarray}}
\def\ea{\end{eqnarray}}

\newcommand{\beqa}{\begin{eqnarray}}
\newcommand{\eeqa}{\end{eqnarray}}

\newcommand{\si}{\sigma}

\newcommand{\pd}{\partial}

\begin{document}
\begin{center}
\vspace{4.cm}

{\bf \large The Conformal Anomaly and the Neutral Currents Sector of the Standard Model\\}

\vspace{1cm}

{\bf Claudio Corian\`{o}, Luigi Delle Rose, Antonio Quintavalle and Mirko Serino}

\vspace{1cm}

{\it Departimento di Fisica, Universit\`{a} del Salento \\
and  INFN-Lecce, Via Arnesano 73100, Lecce, Italy\footnote{claudio.coriano@unisalento.it, luigi.dellerose@le.infn.it, antonio.quintavalle@le.infn.it, mirko.serino@le.infn.it}
}\\
\vspace{.5cm}
\begin{abstract}
We elaborate on the structure of the graviton-gauge-gauge vertex in the electroweak theory, obtained by the insertion of the complete energy-momentum tensor ($T$) on 2-point functions of neutral gauge currents ($VV'$). The vertex defines the leading contribution to the effective action which accounts for the conformal anomaly and related interaction between the Standard Model and gravity. The energy momentum tensor is derived from the curved spacetime Lagrangian in the linearized gravitational limit, and with the inclusion of the term of improvement of a conformally coupled Higgs sector.  As in the previous cases of QED and QCD, we find that the conformal anomaly induces an effective massless scalar interaction between gravity and the neutral currents in each gauge invariant component of the vertex. This is described by the exchange of an anomaly pole. We show that for a spontaneously broken theory 
the anomaly can be entirely attributed to the poles only for a conformally coupled Higgs scalar. In the exchange of a graviton, the trace part of the corresponding interaction can be interpreted as due to an effective dilaton, using a local version of the effective action.  We discuss the implications of the anomalous Ward identity for the $TVV'$ correlator for the structure of the gauge/gauge/ effective dilaton vertex in the effective action.  The analogy between these effective interactions and those related to the radion in theories with large extra dimensions is pointed out. 

\end{abstract}
\end{center}
\newpage

\section{Introduction} 
Gravity couples to the Standard Model, in the weak gravitational field limit, via its energy momentum tensor (EMT) $T^{\mu\nu}$.  This interaction is responsible for the generation of the radiative breaking of scale invariance \cite{Duff:1977ay,Adler:1976zt,Freedman:1974gs}, which is mediated, at leading order in the gauge coupling ($O(g^2)$), by a triangle diagram: the $TVV'$ vertex (see \cite{Giannotti:2008cv,Armillis:2009pq,Armillis:2010pa,Armillis:2010qk}), where $V, V'$ denote two gauge bosons. The computation of the vertex is rather involved, due to the very lengthy expression of the EMT in the electroweak theory, but also not so obvious, due to the need to extract the correct external constraints which are necessary for its consistent definition. 

 The constraints take the form of 3 Ward identities derived by the conservation of the EMT and of (at least) 3 Slavnov-Taylor identities (STI's) on the gauge currents. All of them need to be checked in perturbation theory in a given regularization scheme, in order to secure the consistency of the result.  In the case under exam they correspond to the $TAA, TAZ$ and $TZZ$ vertices, where $A$ is the photon and $Z$ the neutral massive electroweak gauge boson. We will be stating these identities omitting any proof, since the details of the derivations are quite involved.

\begin{figure}[t]
\centering
\subfigure[]{\includegraphics[scale=1.5]{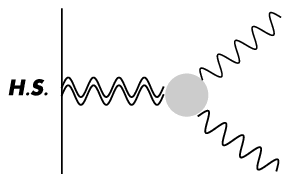}} \hspace{2.cm}
\subfigure[]{\includegraphics[scale=1.5]{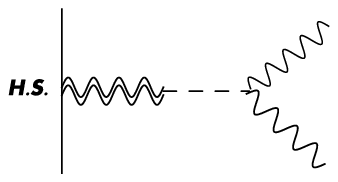}} 
\caption{Gravitational interaction of the Standard Model fields with a hidden sector (H.S.), at leading order in the gravitational constant (a). The interaction in perturbation theory responsible for the trace anomaly is illustrated in (b) via the exchange of an anomaly pole.}
\label{hidden1}
\end{figure}

The explicit computation of these radiative corrections (i.e. of the anomalous action) finds two direct applications. The first has to 
do with the analysis of anomaly mediation as a possible mechanism to describe the interaction between a hypothetical hidden sector and the fields of the Standard Model, 
as shown in Fig. \ref{hidden1} (a). One of the results of our analysis, in this context, is that anomaly mediation is described by the  exchange of anomaly poles in each gauge invariant sector of the perturbative expansion, as shown in Fig. \ref{hidden1} (b). This feature, already present in the QED and QCD cases, as we will comment below, is indeed confirmed by the direct computation in the entire electroweak theory. One of the main implications of our analysis, in fact, is that this picture remains valid even in the presence of mass corrections due to symmetry breaking, for a graviton of large virtuality and a conformally coupled Higgs sector. We will comment on this point in a separate section (section 5) and in our summary before the conclusions.

\begin{figure}[t]
\begin{align}
\begin{minipage}[c]{120pt}
\includegraphics[scale=1.0]{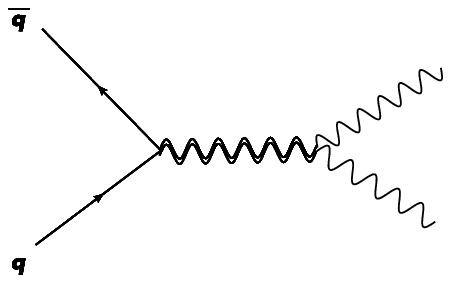}
\end{minipage}
+
\begin{minipage}[c]{150pt}
\includegraphics[scale=1.0]{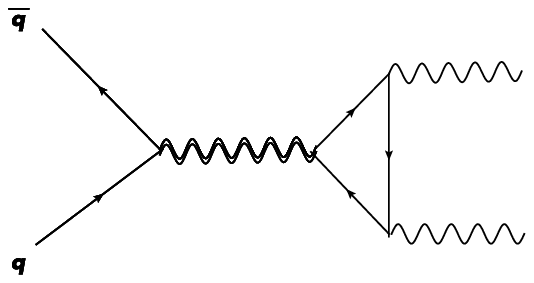}
\end{minipage} 
+ \ldots
\nn
\end{align}
\caption{Typical leading order ($O(\kappa^2)$) contributions to the production of two gauge bosons with gravitational mediation.  
Not included are the initial state (Standard Model) corrections on the $q\bar{q}$/graviton vertex and the loops of gauge bosons and Higgs mediating the decay of the graviton. The latter contribute to the conformal anomaly. }
\label{LHC}
\end{figure}

A second area where these corrections may turn useful is in the case of an electroweak theory formulated in scenarios with large extra dimensions (LED), with a reduced scale for gravity. In this case the virtual exchanges of gravitons provide sizeable corrections to electroweak processes - beyond tree level - useful for LHC studies of these models, as illustrated in Fig. \ref{LHC} in the case of the $q\bar{q}$ annihilation channel. In these extensions a graviscalar  (radion) $\phi$ degree of freedom is induced by the compactification, which is expected to couple to the anomaly ($\phi T^\mu_\mu$) as well as to the scaling-violating terms, as we are going to clarify, by an extra prescription. This prescription is based on the replacement of the classical trace of the matter EMT by its quantum average. A rigorous discussion of the fundamental anomalous 
Ward identity for the $TVV'$ correlator will clarify some subtle issues involved in this prescription. We will show, in parallel, that the anomalous effective action induces in the 1-graviton exchange channel a similar interaction. This interaction can be thought as being mediated by an effective massless dilaton, coupled to the trace anomaly equation (and to its mass corrections).

\section{Definitions and Ward and Slavnov-Taylor identities}
We start with few definitions, focusing our discussions only on the case of the graviton/photon/photon ($TAA$) and graviton/Z/Z 
($TZZ$) vertices. 

We recall that the fundamental action describing the gravity and the Standard Model is defined by the three contributions 
\beq S = S_G + S_{SM} + S_{I}= -\frac{1}{\kappa^2}\int d^4 x \sqrt{-g}\, R + \int d^4 x
\sqrt{-g}\mathcal{L}_{SM} + \frac{1}{6} \int d^4 x \sqrt{-g}\, R \, \mathcal H^\dag \mathcal H      \, ,
\eeq
where $\kappa^2=16 \pi G_N$, with $G_N$ being the four dimensional Newton's constant and $\mathcal H$ is the Higgs doublet. 
We have denoted with $S_G$ the contribution from gravity (Einstein-Hilbert term) while $S_{SM}$ is the Standard Model (SM) quantum action, extended to curved spacetime. $S_I$ denotes the term of improvement for the scalars, which are coupled to the metric via its scalar curvature $R$.
The factor $1/6$ should be recognized as giving a conformally coupled Lorentz scalar, the $SU(2)$ Higgs doublet.
The EMT in our conventions is defined as 
\beq  T_{\mu\nu}(x)  = \frac{2}{\sqrt{-g(x)}}\frac{\d [S_{SM} + S_I ]}{\d g^{\mu\nu}(x)}, 
\eeq
and around a flat spacetime limit 
\beq\label{QMM} g_{\mu\nu}(x) = \h_{\mu\nu} + \kappa \, h_{\mu\nu}(x)\,,\eeq
with the symmetric rank-2 tensor $h_{\mu\nu}(x)$ accounting for the metric fluctuations.

We denote with $T_{\mu\nu}$ the complete (quantum) EMT of the electroweak sector of the Standard Model. This includes the contributions of all the physical fields and of the Goldstones and ghosts in the broken electroweak phase. Its expression is uniquely given by the coupling of the Standard Model Lagrangian to gravity, modulo the terms of improvements, which depend on the choice of the coupling of the Higgs doublets. As we have mentioned, we have chosen a conformally coupled Higgs field.  Our computation is performed in the 
$R_\xi$ gauge.
The expression of the EMT is symmetric and conserved. 
It is therefore given by a minimal contribution $T^{Min}_{\mu\nu}$ (without improvement) and the improvement EMT,  $T^I_{\mu\nu}$, with 
\bea
T_{\mu\nu} = T^{Min}_{\mu\nu} + T^I_{\mu\nu} \,,
\eea
where the minimal tensor is decomposed into
\bea
T^{Min}_{\mu\nu} = T^{f.s.}_{\mu\nu} + T^{ferm.}_{\mu\nu} + T^{Higgs}_{\mu\nu} + T^{Yukawa}_{\mu\nu} + T^{g.fix.}_{\mu\nu} + T^{ghost}_{\mu\nu}.
\eea
The various contributions refer, respectively, to the gauge kinetic terms (field strength, $f.s.$), the fermions, the Higgs, Yukawa, gauge fixing contributions ($g.fix.$) and the contributions coming from the ghost sector. 
As we have already mentioned, in order to fix the structure of the correlator one needs to derive and implement the necessary Ward and STI's. Their derivation is quite lengthy as is their implementation in perturbation theory, given the sizeable number of diagrams involved in the expansion and the very long expression of the vertex extracted from the EMT.

We obtain:\\ 
1) A Ward identity related to the conservation of the EMT in the flat spacetime limit (i.e. $\partial_{\mu} T^{\mu\nu}=0$), which takes the form 
\beqa
\label{WardTVV}
 -i\frac{\kappa}{2}\pd^{\mu}\langle T_{\mu\nu}(x) V_\a (x_1) V'_{\b} (x_2)\rangle_{amp}
&=& - \frac{\kappa}{2}\bigg\{- \pd_\nu\d^{(4)}(x_1-x) P^{-1\,VV'}_{\a\b}(x_2,x) \nn \\
&& \hspace{-6cm} - \pd_\nu\d^{(4)}(x_2-x) P^{-1\,V V'}_{\a\b}(x_1,x)
+ \pd^\mu[ \h_{\a\nu} \d^{(4)}(x_1-x) P^{-1\,V V'}_{\b\mu}(x_2,x) + \h_{\b\nu}\delta^{(4)}(x_2-x) P^{-1\,V V'}_{\a\mu}(x_1,x)]\bigg\}, \nn 
\eeqa
where we have introduced the off-diagonal 2-point function
\beq P^{-1\,VV'}_{\a\b}(x_1,x_2) = \langle 0| T V_{\a}(x_1) V'_{\b}(x_2) | 0 \rangle_{amp},
\eeq
where $amp$ denotes amputated external gauge lines. Notice that the gravitational field, in this computation, is just an external 
field and the 1PI conditions apply only to the external gauge lines. This point emerges from a closer investigation of the defining STI's of the correlator. This Ward identity applies to any gauge boson in the neutral sector. 

2) A STI for the $TAA$ vertex. Specifically, introducing the photon gauge-fixing function
\beqa
\mathcal F^A  =  \pd^\s A_\s\, ,\qquad \qquad
\eeqa
we obtain the relation  
\beqa
 \frac{1}{\xi}\langle T_{\mu\nu}(z)\mathcal F^A(x)\mathcal F^A(y) \rangle &=& 
- \frac{i}{\xi}\bigg\{ \eta_{\mu\nu} \, \pd^\r_x \left[\d^{(4)}(z - x)  \langle A_\r(x) \mathcal F^A(y) \rangle \right] \nn \\
&& \hspace{-4cm} - \eta_{\mu\nu}\pd^\r_z  \left[ \d^{(4)}(z-y) \langle A_\r(z) \mathcal F^A(x) \rangle \right] 
 - \bigg( \pd_\mu^x \left[ \d^{(4)}(z - x)  \langle A_\nu(x) \mathcal F^A(y)\rangle \right]  \nn \\
&& \hspace{-4cm} -  \pd_\mu^z \d^{(4)}(z-y)  \langle A_\nu(z) \mathcal F^A(x) \rangle  + (\mu \leftrightarrow \nu)  \bigg)   \bigg\} \, ,
\eeqa
with $\xi$ denoting the gauge-fixing parameter. \\
\begin{figure}[t]
\centering
\subfigure[]{\includegraphics[scale=0.8]{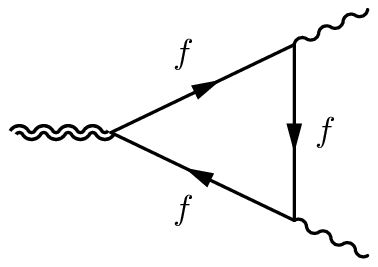}} \hspace{.5cm}
\subfigure[]{\includegraphics[scale=0.8]{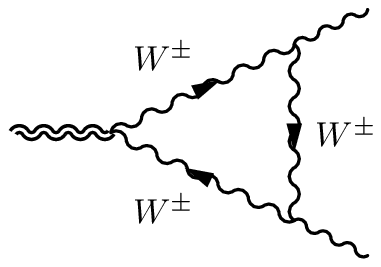}} \hspace{.5cm}
\subfigure[]{\includegraphics[scale=0.8]{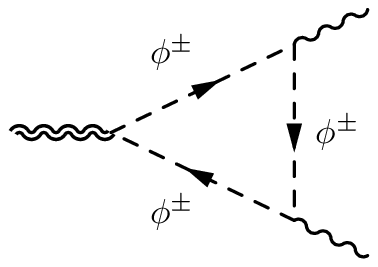}} \hspace{.5cm}
\subfigure[]{\includegraphics[scale=0.8]{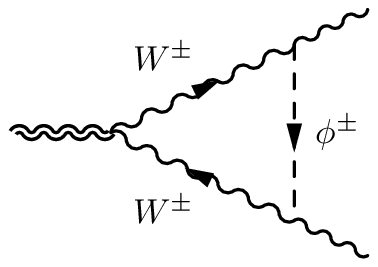}}
\\
\vspace{.5cm}
\subfigure[]{\includegraphics[scale=0.8]{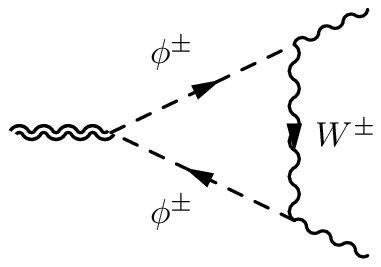}} \hspace{.5cm}
\subfigure[]{\includegraphics[scale=0.8]{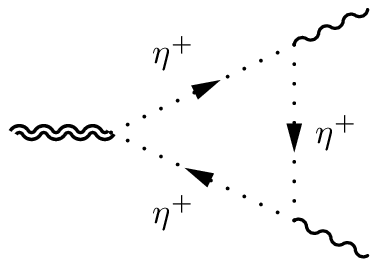}} \hspace{.5cm}
\subfigure[]{\includegraphics[scale=0.8]{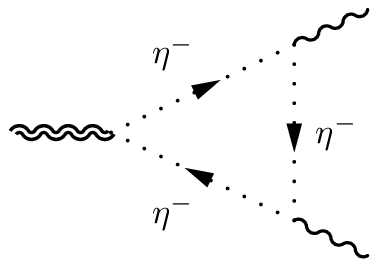}} \hspace{.5cm}
\caption{Amplitudes with the triangle topology for the two correlators $TAA$ and $TZZ$.  \label{triangles}}
\end{figure}
\begin{figure}[t]
\centering
\subfigure[]{\includegraphics[scale=0.75]{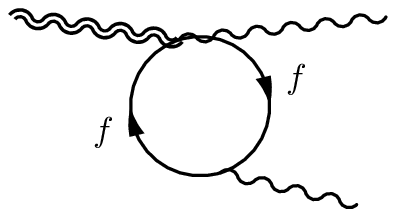}}\hspace{.5cm}
\subfigure[]{\includegraphics[scale=0.75]{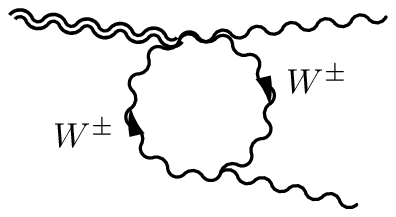}}\hspace{.5cm}
\subfigure[]{\includegraphics[scale=0.75]{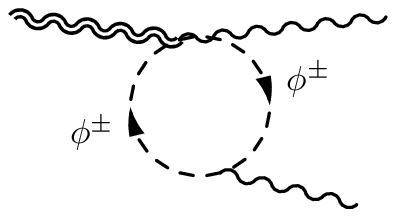}}\hspace{.5cm}
\subfigure[]{\includegraphics[scale=0.75]{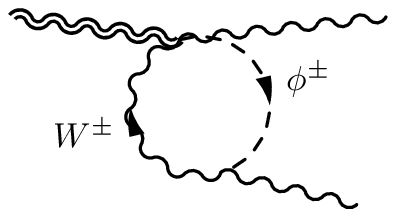}}
\\
\vspace{.5cm}
\subfigure[]{\includegraphics[scale=0.75]{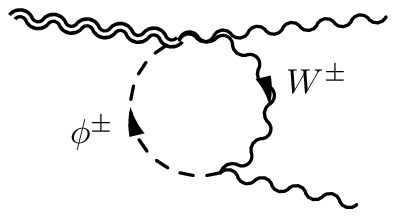}}\hspace{.5cm}
\subfigure[]{\includegraphics[scale=0.75]{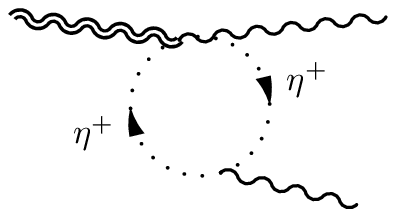}}\hspace{.5cm}
\subfigure[]{\includegraphics[scale=0.75]{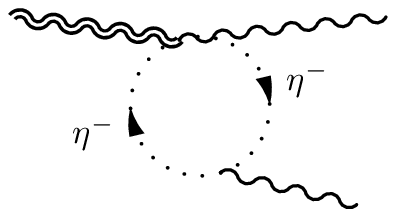}}\hspace{.5cm}
\caption{Amplitudes with t-bubble topology for the correlators $TAA$ and $TZZ$. \label{t-bubble}}
\end{figure}
\begin{figure}[t]
\centering
\subfigure[]{\includegraphics[scale=0.75]{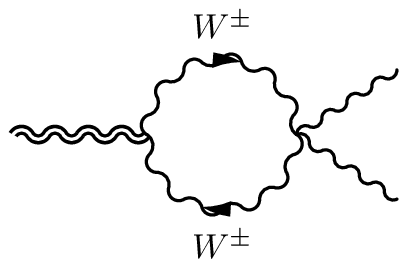}}\hspace{.5cm}
\subfigure[]{\includegraphics[scale=0.75]{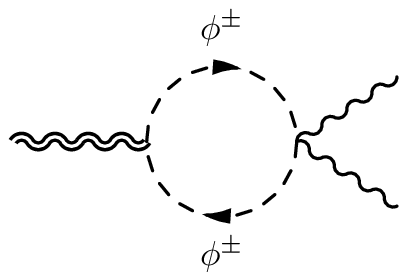}}
\caption{Amplitudes with s-bubble topology for the correlators $TAA$ and $TZZ$. \label{s-bubble}}
\end{figure}
\begin{figure}[t]
\centering
\subfigure[]{\includegraphics[scale=0.8]{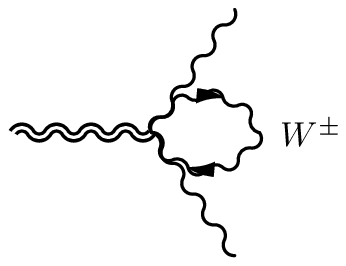}}\hspace{.5cm}
\subfigure[]{\includegraphics[scale=0.8]{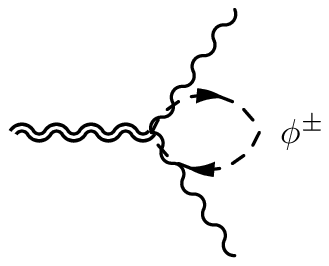}}
\caption{Amplitudes with the tadpole topology for the correlators $TAA$ and $TZZ$.\label{tadpoles}}
\end{figure}
\begin{figure}[t]
\centering
\subfigure[]{\includegraphics[scale=0.8]{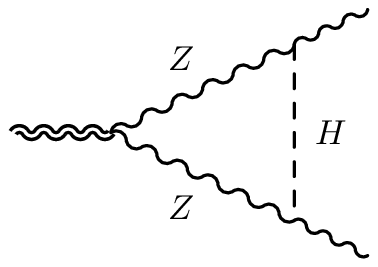}}\hspace{.5cm}
\subfigure[]{\includegraphics[scale=0.8]{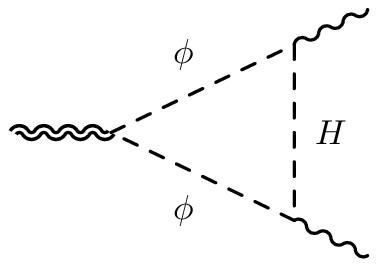}}\hspace{.5cm}
\subfigure[]{\includegraphics[scale=0.8]{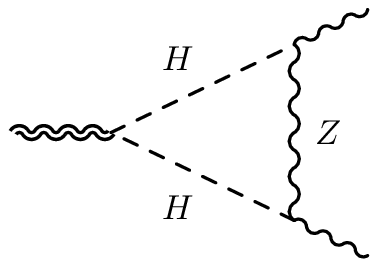}}\hspace{.5cm}
\subfigure[]{\includegraphics[scale=0.8]{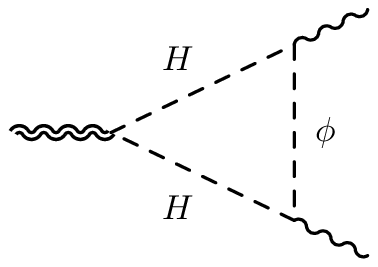}}
\caption{Amplitudes with the triangle topology for the correlator $TZZ$. \label{triangles1}}
\end{figure}
\begin{figure}[t]
\centering
\subfigure[]{\includegraphics[scale=0.75]{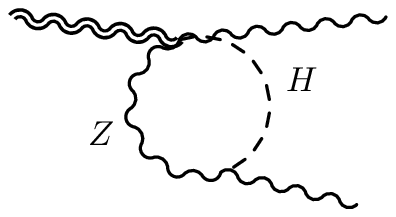}}\hspace{.5cm}
\subfigure[]{\includegraphics[scale=0.75]{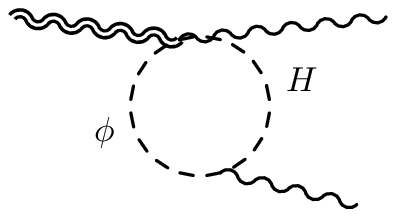}}
\caption{Amplitudes with the t-bubble topology for the correlator $TZZ$.\label{t-bubble1}}
\end{figure}
\begin{figure}[t]
\centering
\subfigure[]{\includegraphics[scale=0.75]{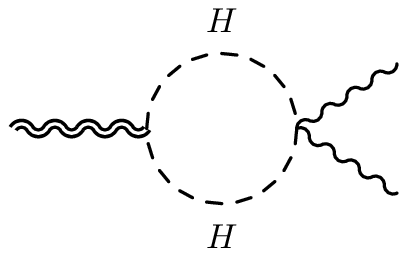}}\hspace{.5cm}
\subfigure[]{\includegraphics[scale=0.75]{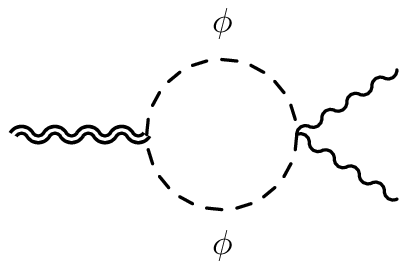}}
\caption{Amplitudes with the s-bubble topology for the correlator $TZZ$.\label{s-bubble1}}
\end{figure}
\begin{figure}[t]
\centering
\subfigure[]{\includegraphics[scale=0.8]{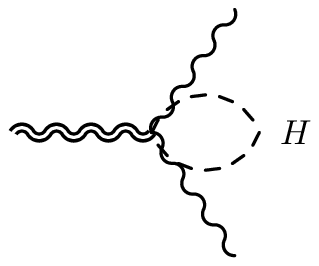}}\hspace{.5cm}
\subfigure[]{\includegraphics[scale=0.8]{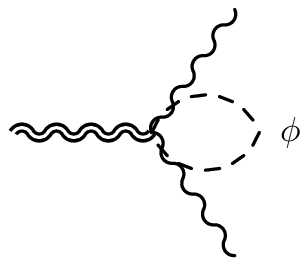}}\hspace{.5cm}
\subfigure[]{\includegraphics[scale=0.8]{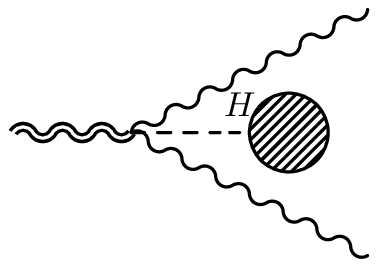}}\hspace{.5cm}
\caption{Amplitudes with tadpole topology for the correlator $TZZ$.\label{tadpoles1}}
\end{figure}
3) A STI for the $TZZ$ correlator. Introducing the gauge-fixing function of the $Z$ gauge boson 
\beqa
\mathcal F^Z  =  \pd^\s Z_\s - \xi M_Z \f\, ,
\eeqa
where  $\phi$ is the Goldstone of the $Z$, this takes the form 
\beqa\label{STZZfinalcoord}
\frac{1}{\xi}\langle T_{\mu\nu}(z)\mathcal F^Z(x)\mathcal F^Z(y)\rangle
&=& -\frac{i}{\xi}\bigg\{-i\xi^2\h_{\mu\nu}\d^{(4)}(x-y)\d^{(4)}(x-z) + \h_{\mu\nu}\pd^\r_x\bigg[\d^{(4)}(x-z)\bigg]\langle Z_\r(x)\mathcal F^Z(y)\rangle\nn\\
&-&  \pd_\mu^x\bigg[\d^{(4)}(x-z)\langle Z_\nu(x)\mathcal F^Z(y)\rangle\bigg] - \pd_\nu^x\bigg[\d^{(4)}(x-z)\langle Z_\mu(x)\mathcal F^Z(y)\rangle\bigg]\nn\\
&+&  \pd_\mu^z\bigg[\d^{(4)}(z-y)\bigg]\langle Z_\nu(z)\mathcal F^Z(x)\rangle + \pd_\nu^z\bigg[\d^{(4)}(z-y)\bigg]\langle Z_\mu(z)\mathcal F^Z(x)\rangle\nn\\
&-&  \h_{\mu\nu}\pd^\r_z\bigg(\d^{(4)}(z-y)\langle Z_\r(z)\mathcal F^Z(x)\rangle\bigg)\bigg\} \, .
\eeqa
4) A STI for the $TAZ$ vertex
\beqa
\frac{1}{\xi}\langle T_{\mu\nu}(z) \mathcal F^A(x)\mathcal F^Z(y)\rangle
&=& - \frac{i}{\xi}\bigg\{-\h_{\mu\nu}\pd^\s_z\bigg[\d^{(4)}(z-y)\langle
       Z_\s(z) \mathcal F^A(x)\rangle\bigg] \nn\\
&+& \pd_\nu^z\d^{(4)}(z-y) \langle Z_\mu(z)\mathcal F^A(x)\rangle + \pd_\mu^z\d^{(4)}(z-y) \langle Z_\nu(z)\mathcal F^A(x)\rangle\bigg\}\, . 
\eeqa

We illustrate the overall structure of the results for the $TAA$ and $TZZ$  vertices, focusing on the essential parts, and in particular on those form factors which contribute to the trace part, since they are simpler. The complete result is indeed quite involved and some details can be found in \cite{Coriano:2011zk}.

\section{Results for the $TAA$ case}
In the $TAA$ case, we introduce the notation $\Gamma^{(AA)\mu\nu\alpha\beta}(p,q)$ to denote the one-loop amputated vertex function with a graviton and two on-shell photons.

In momentum space we indicate with $k$ the momentum of the incoming graviton and with 
$p$ and $q$ the momenta of the two photons. In general, the $\Gamma^{(VV')\mu\nu\alpha\beta}(p,q)$ correlator 
is defined as
\beqa
(2\pi)^4\d^{(4)}(k-p-q)\Gamma^{VV'}_{\mu\nu\a\b}(p,q)&=&
 -i\frac{\kappa}{2}\int d^4zd^4xd^4y\, \langle T_{\mu\nu}(z) V_{\a}(x) V'_{\b}(y)\rangle_{amp}\,
e^{-ikz + ipx + iqy}.\, \eeqa

In the 2-photon case $(AA)$ is decomposed in the form

\bea
\Gamma^{(AA)\mu\nu\a\b}(p,q) = \Gamma_{F}^{(AA)\mu\nu\a\b}(p,q) + \Gamma_{B}^{(AA)\mu\nu\a\b}(p,q) + \Gamma_{I}^{(AA)\mu\nu\a\b}(p,q),
\eea
as a sum of a fermion sector (F) (Fig. \ref{triangles}(a), Fig. \ref{t-bubble}(a)), a gauge boson sector (B) (Fig. \ref{triangles}(b)-(g), Fig. \ref{t-bubble}(b)-(g), Fig. \ref{s-bubble}, Fig. \ref{tadpoles}) and a term of improvement denoted as $\Gamma^{\mu\nu\a\b}_{I}$. The contributions to the (F) and (B) sectors are obtained by the insertion of $T^{Min}$. The contribution from the term of improvement is given by diagrams of the same form of those in Fig. \ref{triangles}(c), 3(e) and Fig. \ref{s-bubble}(b), but now with the graviton - scalar - scalar vertices determined only by the energy momentum tensor $T_I^{\mu\nu}$. 

The tensor basis on which we expand the vertex is given by four independent tensor structures
\bea
\label{phis}
  \phi_1^{\, \mu \nu \a \b} (p,q) &=&
 (s \, \eta^{\mu\nu} - k^{\mu}k^{\nu}) \, u^{\a \b} (p,q),
 \label{widetilde1}\\
\phi_2^{\, \mu \nu \a \b} (p,q) &=& - 2 \, u^{\a \b} (p,q) \left[ s \, \eta^{\mu \nu} + 2 (p^\mu \, p^\nu + q^\mu \, q^\nu )
- 4 \, (p^\mu \, q^\nu + q^\mu \, p^\nu) \right],
\label{widetilde2} \\
\phi^{\, \mu \nu \alpha \beta}_{3} (p,q) &=&
\big(p^{\mu} q^{\nu} + p^{\nu} q^{\mu}\big)\eta^{\alpha\beta}
+ \frac{s}{2} \left(\eta^{\alpha\nu} \eta^{\beta\mu} + \eta^{\alpha\mu} \eta^{\beta\nu}\right) \nn \\
&&  \hspace{1cm} - \eta^{\mu\nu} \left(\frac{s}{2} \eta^{\alpha \beta}- q^{\alpha} p^{\beta}\right)
-\left(\eta^{\beta\nu} p^{\mu}
+ \eta^{\beta\mu} p^{\nu}\right)q^{\alpha}
 - \big (\eta^{\alpha\nu} q^{\mu}
+ \eta^{\alpha\mu} q^{\nu }\big)p^{\beta} \nn \\
\phi_4^{\mu\nu\alpha\beta}(p,q) &=& (s \, \eta^{\mu\nu} - k^{\mu}k^{\nu}) \, \eta^{\alpha\beta} 
\label{widetilde3}
\eea
where $u^{\a \b} (p,q)$ has been defined as
\beq
u^{\alpha\beta}(p,q) \equiv (p\cdot q) \,  \eta^{\alpha\beta} - q^{\alpha} \, p^{\beta},\,\\
\label{utensor}
\eeq
among which only  $\phi_1^{\, \mu \nu \a \b}$ and $\phi_4^{\, \mu \nu \a \b}$ show manifestly a trace, the remaining ones being traceless. A complete computation gives for the various gauge invariant subsectors 
\bea
\Gamma^{(AA)\mu\nu\alpha\beta}_{F}(p,q) &=&  \, \sum_{i=1}^{3} \Phi_{i\,F} (s,0, 0,m_f^2) \, \phi_i^{\mu\nu\alpha\beta}(p,q)\,, \\
\Gamma^{(AA)\mu\nu\alpha\beta}_{B}(p,q) &=&  \, \sum_{i=1}^{3} \Phi_{i\,B} (s,0, 0,M_W^2) \, \phi_i^{\mu\nu\alpha\beta}(p,q)\,, \\
\Gamma^{(AA)\mu\nu\alpha\beta}_{I}(p,q) &=&  \Phi_{1\,I} (s,0, 0,M_W^2) \, \phi_1^{\mu\nu\alpha\beta}(p,q) + \Phi_{4\,I} (s,0, 0,M_W^2) \, \phi_4^{\mu\nu\alpha\beta}(p,q) \,.
\eea
  The first three arguments of the form factors stand for the three independent kinematical invariants $k^2 = (p+q)^2 = s$, $p^2 = q^2 = 0$ while the remaining ones denote the particle masses circulating in the loop. We use the on-shell renormalization scheme.

As already shown in the QED and QCD cases \cite{Armillis:2009pq,Armillis:2010pa,Armillis:2010qk}, in an unbroken gauge theory the entire contribution to the trace anomaly comes from the first tensor structure $\phi_1$. 

In the $TAA$ vertex, the contribution to the trace anomaly in the fermion sector comes from $\Phi_{1\, F}$ which is given by 
\bea
\Phi_{1\, F} (s,\,0,\,0,\,m_f^2) &=& - i \frac{\kappa}{2}\, \frac{\alpha}{3 \pi \, s} \sum_{f} Q_f^2 \bigg\{
- \frac{2}{3} + \frac{4\,m_f^2}{s} - 2\,m_f^2 \, \mathcal C_0 (s, 0, 0, m_f^2, m_f^2, m_f^2)
\bigg[1 - \frac{4 m_f^2}{s}\bigg] \bigg\}.  
\label{taavertex}
\eea
Here we have introduced the QED coupling  $\alpha=e^2/(4 \pi)$ and the function
\beq \mathcal C_0 (s,0,0,m^2,m^2,m^2) = \frac{1}{2 s} \log^2 \frac{a_3+1}{a_3-1}\, \eeq
obtained from the scalar triangle integral, with
\bea
a_3 = \sqrt {1- \frac{4 m^2}{s}}.\qquad \qquad
\eea
The sum is taken over all the fermions ($f$) of the Standard Model. 
As one can immediately realize, this form factor is characterized by the presence of an anomaly pole
\beq
\Phi^F_{1\, pole}\equiv i \kappa \frac{\alpha}{9 \pi \, s} \sum_{f} Q_f^2
\eeq
which is responsible for the generation of the anomaly in the massless limit. To appreciate the significance of this "pole contribution" one needs special care, since a computation of the residue (at $s=0$) shows that this is indeed zero in the presence of mass corrections.
However, this leading $1/s$ behaviour in the trace part of the amplitude, as we are going to show, is clearly identifiable in  an (asymptotic) expansion ($s \gg m_f^2$), and is corrected by extra $m_f^2/s^2$ terms, where $m_f$ denotes generically any fermion of the SM. In other words, this component is extracted in the UV limit of the amplitude even in the massive case and is a clear manifestation of the anomaly.

The other gauge-invariant sector of the $TAA$ vertex is the one mediated by the exchange of bosons, Goldstones and ghosts in the loop. We will denote with $M_W,M_Z$ and $M_H$ the masses of the W's and $Z$ gauge bosons and the Higgs mass respectively.
In this
sector the form factor contributing to the trace is
\bea
\Phi_{1\, B} (s,\,0,\,0,\,M_W^2) &=& - i \frac{\kappa}{2}\, \frac{\alpha}{\pi \, s} \bigg\{
\frac{5}{6} - \frac{2\,M_W^2}{s} + 2\,M_W^2 \, \mathcal C_0 (s, 0, 0, M_W^2, M_W^2, M_W^2)
\bigg[1 - \frac{2 M_W^2}{s}\bigg] \bigg\}, 
\label{oneb} 
\eeqa
which multiplies the tensor structure $\phi_1$, responsible for the generation of the anomalous trace. 

In this case the anomaly pole is easily isolated from (\ref{oneb}) in the form
\beq \label{Phi1Bpole}
\Phi_{1\, B, pole}\equiv - i \frac{\kappa}{2}\, \frac{\alpha}{\pi \, s} \frac{5}{6} \,.
\eeq
The term of improvement is responsible for the generation of two form factors, both of them contributing to the trace. They are given by 
\bea
\Phi_{1\, I} (s,\,0,\,0,\,M_W^2) &=& - i \frac{\kappa}{2}\frac{\alpha}{3 \pi \, s} \bigg\{ 1 + 2 M_W^2 \,C_0 (s, 0, 0, M_W^2, M_W^2, M_W^2)\bigg\} ,\\
\Phi_{4\, I} (s,\,0,\,0,\,M_W^2) &=&  i \frac{\kappa}{2}\frac{\alpha}{6 \pi }  M_W^2 \,C_0 (s, 0, 0, M_W^2, M_W^2, M_W^2), 
\eea
the first of them being characterized by an anomaly pole 
\beq \label{Phi1Ipole}
\Phi_{1\, I\,\, pole}=  - i \frac{\kappa}{2}\frac{\alpha}{3 \pi \, s}.
\eeq
Our considerations on the UV behaviour of the  a) {\em radiative} plus the b) {\em explicit mass corrections} to the anomalous amplitude are obviously based on an exact computation of the correlator. 

In the asymptotic limit ($s\to \infty$), the expansions of the three form factors contributing to the trace part can be organized in terms of the $1/s$ "pole component" plus mass corrections, which are given by 
\beqa
\Phi_{1,F}(s,0,0,m_f^2) &\simeq& -i\frac{\kappa}{2} \frac{\alpha}{3\pi s} \sum_{f} Q_f^2 \bigg\{-\frac{2}{3} + \frac{m_f^2}{s}\bigg[ 4 + \pi^2
-\log^2(\frac{m_f^2}{s})-2i\pi\log(\frac{m_f^2}{s})\bigg]\bigg\}\,,\\
\Phi_{1,B}(s,0,0,M_W^2)&\simeq&-i\frac{\kappa}{2}\frac{\alpha}{\pi s}\bigg\{\frac{5}{6}-\frac{M_W^2}{s}\bigg[2+\pi^2
-\log^2(\frac{M_W^2}{s})-2i\pi\log(\frac{M_W^2}{s})\bigg]\bigg\}\,,\\
\Phi_{1,I}(s,0,0,M_W^2)& \simeq& -i\frac{\kappa}{2}\frac{\alpha}{3\pi s}\bigg\{1-\frac{M_W^2}{s}\bigg[\pi - i \log(\frac{M_W^2}{s})
\bigg]^2\bigg\} \,, 
\label{expan}
\eeqa
The energy suppressed terms ($m_f^2/s^2$, $M_W^2/s^2$) take the typical form $M^2/s^2$, with $M$ 
denoting, generically, any explicit mass term generated in the broken phase of the theory.  This separation of the radiative from the 
explicit contributions to the breaking of conformal invariance, due to the tree-level mass terms, is in agreement with the obvious fact that in the UV limit,  masses can be dropped. At the same time the (radiative) breaking of the conformal symmetry remains, with no much surprise.   

Preliminarily, we recall that in the $\overline{MS}$ scheme the $\beta$ functions of the Standard Model are given by
\bea
\label{betas}
\beta_1 = \frac{g_1^3}{16 \pi^2} \bigg[ \frac{20}{9} n_g + \frac{1}{6} \bigg] \,, \qquad
\beta_2 = \frac{g_2^3}{16 \pi^2} \bigg[ \frac{4}{3} n_g - \frac{22}{3} + \frac{1}{6} \bigg] \,, \qquad
\beta_3 = \frac{g_3^3}{16 \pi^2} \bigg[ - 11 + \frac{4}{3} n_g \bigg] \,, 
\eea
for the hypercharge, weak and strong interactions respectively, and $n_g$ is the number of generations. 
The expression of the $\beta$ function of the electromagnetic coupling, $\beta_e$, is given in the same scheme by
\bea
\beta_e = c_w^2 \, \beta_1 + s_w^2 \, \beta_2 = \frac{e^3}{16 \pi^2} \bigg[ \frac{32}{9} n_g - 7 \bigg].
\eea

At this point, the residue of the anomaly pole which appears in the form factors $\Phi_{1,F}$, $\Phi_{1,B}$ and $\Phi_{1,I}$ is uniquely determined by the beta function of the electromagnetic coupling constant. Indeed we have
\bea
\Phi_{1, pole} = \Phi_{1, pole}^F + \Phi_{1, pole}^B + \Phi_{1, pole}^I = - i \frac{\kappa}{2} \frac{\alpha}{3 \pi s} \bigg[ - \frac{2}{3} \sum_f Q_f^2 + \frac{5}{2}+ 1\bigg] =  i  \frac{\kappa}{3 s} \frac{\beta_e}{e} \,,
\eea
where we have used the fact that $\sum_f Q_f^2 = \frac{8}{3} n_g $.

\section{Results for the $TZZ$ case}
Moving to the vertex with two massive $Z$ gauge bosons, one discovers a similar pattern. 
Also in this case, as before, we introduce the notation $\Gamma^{(ZZ)\mu\nu\alpha\beta}(p,q)$
to describe the corresponding correlation function. We have several contributions appearing in the global expression of the correlator:
\bea
\Gamma ^{(ZZ)\mu\nu\alpha\beta}(p,q) = \Gamma ^{(ZZ)\mu\nu\alpha\beta}_F (p,q)  + \Gamma^{(ZZ)\mu\nu\alpha\beta}_W (p,q) + \Gamma^{(ZZ)\mu\nu\alpha\beta}_{Z,H} (p,q) +  \Gamma^{(ZZ)\mu\nu\alpha\beta}_I (p,q).
\eea
$\Gamma^{(ZZ)\mu\nu\alpha\beta}(p,q)$, for on-shell $Z$ bosons, can be separated into three contributions obtained using 
the insertion of $T^{Min}$ (sectors $F, W$ and $Z/H$) and a fourth one coming from the term of improvement. In this case the gravitational interaction is mediated by $T_I$.

The labelling of the first three is inherited from the types of particles (and corresponding masses) that circulate in the loops. Beside the fermion sector (F) with diagrams depicted in Figs. \ref{triangles}(a) and \ref{t-bubble}(a)), the other contributions involve a $W$ gauge boson (sector (W)), with diagrams Fig. \ref{triangles}(b)-(g), Fig. \ref{t-bubble}(b)-(g), Fig. \ref{s-bubble} and Fig. \ref{tadpoles}), and the mixed $Z$/Higgs bosons sector $(Z,H)$ with contributions shown in Figs. \ref{triangles1}, \ref{t-bubble1}, \ref{s-bubble1} and \ref{tadpoles1}). There is also a diagram proportional to a Higgs tadpole (Fig. \ref{tadpoles1}(a)) which vanishes in the on-shell renormalization scheme. Finally there is a contribution from the term of improvement (I). This is given by the diagrams depicted in Fig. \ref{triangles}(c), (d), \ref{s-bubble}(b), together with those of Figs. \ref{triangles1}(b), (c), (d) and Fig. \ref{s-bubble1}. In this case, however, the graviton - scalar - scalar vertices is generated by $T_I^{\mu\nu}$.

As we have already mentioned, we take the two $Z$ gauge bosons on the external lines on-shell, and an insertion of $T_{\mu\nu}$ at a nonzero momentum transfer $k$. The four contributions can be expanded on a tensor basis given by 9 tensors, and corresponding  form factors $\Phi_i$ as
\bea
\Gamma^{(ZZ)\mu\nu\alpha\beta}_F (p,q) &=& \sum_{i = 1}^9 {\Phi^{(F)}_i (s,M_Z^2,M_Z^2,m_f^2) \, t_i^{\mu\nu\alpha\beta}(p,q)} \,, \\
\Gamma^{(ZZ)\mu\nu\alpha\beta}_W (p,q) &=& \sum_{i = 1}^9 {\Phi^{(W)}_i (s,M_Z^2,M_Z^2,M_W^2) \, t_i^{\mu\nu\alpha\beta}(p,q)} \,, \\
\Gamma^{(ZZ)\mu\nu\alpha\beta}_{Z,H} (p,q) &=& \sum_{i = 1}^9 {\Phi^{(Z,H)}_i (s,M_Z^2,M_Z^2,M_Z^2,M_H^2) \, t_i^{\mu\nu\alpha\beta}(p,q)} \,, 
\eea
\bea
\Gamma^{(ZZ)\mu\nu\alpha\beta}_{I} (p,q) &=& \Phi^{(I)}_1 (s,M_Z^2,M_Z^2,M_W^2,M_Z^2,M_H^2) \, t_1^{\mu\nu\alpha\beta}(p,q) \nn \\
&+&  \Phi^{(I)}_2 (s,M_Z^2,M_Z^2,M_W^2,M_Z^2,M_H^2) \, t_2^{\mu\nu\alpha\beta}(p,q) \,, 
\eea
where the first three arguments of the $\Phi_i$'s are the virtualities of the external lines $k^2 = s, \, p^2 = q^2 = M_Z^2$, while the last two give the masses in the internal lines. 
7 of the 9 tensor structures are traceless, while the only two responsible for the breaking of scale invariance are 
\bea
t_1^{\mu\nu\alpha\beta}(p,q) &=&  (s g^{\mu\nu} - k^{\mu}k^{\nu}) \left[ \left( \frac{s}{2}-M_Z^2 \right)g^{\alpha\beta} - q^\alpha p^\beta \right] \,, \nn \\
t_2^{\mu\nu\alpha\beta}(p,q) &=&  (s g^{\mu\nu} - k^{\mu}k^{\nu}) g^{\alpha\beta} \,.
\eea
The four form factors responsible for generating a pole term are those accompanying the tensor structure $t_1$, while the form factors $\Phi_{2}$, corresponding to the tensor structure $t_2$, show no pole. The latter give contributions which are suppressed as $M^2/s^2$. Therefore, as for the $TAA$ vertex, the trace parts show a distinctive $1/s$ contribution plus corrections of $O(M^2/s^2)$, as we have specified above.  Being the complete result of this vertex quite lengthy, we omit details and just focus our attention on the pole terms extracted from each sector. These are summarized by rather simple expressions. We obtain 
\beq
\Phi^{(F)}_{1\, pole}\equiv\frac{i \alpha \, \kappa }{36 \pi c_w^2 s_w^2 \, s}\sum_{f}\left(C_a^{f \, 2}+C_v^{f \, 2}\right)
\eeq
for the fermion sector of the vertex, where $s_w$ and $c_w$ are short notations for $\sin\theta_W$ and $\cos\theta_W$. Similarly, 
in the other sectors we have 
\beq
\Phi^{(W)}_{1\, pole} \equiv  - i \frac{\kappa}{2} \frac{\alpha}{s_w^2 \, c_w^2 \, \pi \, s} \frac{(60s_w^4-148s_w^2+81)}{72}
\eeq
for diagrams involving  $W$'s, while the diagrams with $W$ and $Z$ gauge bosons give
\beq
\Phi^{(Z,H)}_{1\, pole} \equiv \frac{7 i \alpha  \kappa }{144 \pi  s c_w^2 s_w^2}.
\eeq
The term of improvement contributes to two tensor structures but only one of the two form factors from this sector has a pole term. We have, in this case  
 \beq
 \Phi^{(I)}_{1\, pole}  = - i \frac{\kappa}{2} \frac{\alpha}{6 \pi \, s_w^2 \, c_w^2 \, s } \left(1 - 2 s_w^2 c_w^2\right),
\eeq
while  $\Phi^{(I)}_{4}\sim M_W^2/s^2$ asymptotically.

The coefficient of the anomalous pole contribution is fixed by the beta functions of the theory. In this case it is proportional to a linear combination of the beta functions of the couplings $g_1$ and $g_2$ of hypercharge and $SU(2)$. Indeed we have  
\bea
\Phi_{1, pole} = \Phi_{1, pole}^{(F)} + \Phi_{1, pole}^{(W)} + \Phi_{1, pole}^{(Z,H)} + \Phi_{1, pole}^{(I)} =  i \frac{\kappa}{3 s}\bigg[ s_w^2 \frac{\beta_1}{g_1} + c_w^2 \frac{\beta_2}{g_2}\bigg] \,.
\eea
\section{The coupling of the radion/dilaton beyond tree level and the effective dilaton}

One of the most significant applications of the results of the previous sections concerns the study of the coupling of the radion/dilaton to the fields of the Standard Model in theories with LED. We will use the term radion ($\phi$) to denote the fundamental scalar introduced in the usual compactifications of theories with LED, and reserve the name of "effective dilaton" 
($\varphi$) for the scalar interaction dynamically induced by the anomaly.  As we are going to show, the radion has interactions with matter which are quite similar to those allowed to the effective dilaton, although the latter shows up in a different channel (the 1-graviton exchange channel). We proceed first with a rigorous discussion of the interaction of the LED radion and then illustrate the analogies between the two states to clarify these points. 

We recall that in models with LED, with matter on the brane and gravity in the bulk, the compactification of the extra dimensions gives rise in the 4 dimensional effective field theory to towers of Kaluza-Klein gravitons and dilatons. For definiteness we consider a theory compactified on a torus and consider the zero modes of the 4D graviton field and of the dilaton $\phi$ generated by this procedure. These two fields will couple, via their lowest 
Kaluza-Klein modes, to the EMT with the interaction Lagrangian \cite{Han:1998sg}
\bea
\mathcal L_{int} = -\frac{\kappa}{2} \int d^4 x \left( h_{\mu\nu} T^{\mu\nu} + \omega \phi \, T^{\mu}_{\mu} \right) \,, \qquad \omega = \sqrt{\frac{2}{3(\delta+2)}} \, ,
\label{comp}
\eea
where $\delta$ is the number of extra dimensions. To understand the main features of the dilaton interaction at 1-loop level we proceed as follows. \\
We first recall that the structure of the anomaly equation in the presence of a classical trace in a certain theory takes the form 
\bea
\label{traceid}
\eta^{\mu\nu} \langle T_{\mu\nu}(z) \rangle = \langle \mathcal A(z) \rangle + \langle T^{\mu}_{\mu}(z) \rangle\, , 
\eea
where we have taken the quantum average of each term. $\mathcal A$ is the operator describing the anomalous behavior of the fields under scale transformations while the $T^{\mu}_{\mu}$ operator is the non-amomalous contribution to the trace of the EMT. This second term vanishes in the conformal limit (i.e. before electroweak symmetry breaking) using the equations of motion of the fields. In an exact gauge theory the expected structure of the anomaly is given by the relation  
\bea \label{anom}
\mathcal A = \sum_i \frac{\beta_i}{2 g_i} \, F^{\alpha\beta}_i F^i_{\alpha\beta}\, ,
\eea
where $F_i^{\alpha\beta}$ and $g_i$ are the field strengths and the gauge couplings of the gauge fields in the unbroken phase, corresponding to the Standard Model gauge group $SU(3)_C \times SU(2)_L \times U(1)_Y$. For a theory in a broken phase, and in the photon case, the anomaly $\mathcal A$ is again proportional to $\beta_e$.
By taking two functional derivatives of the trace identity (\ref{traceid}) with respect to the sources $J_{\alpha}$ and $J'_{\beta}$ of the gauge fields $V_{\alpha}$ and $V'_{\beta}$, we obtain the anomalous identities on the correlation functions analyzed in this work
\bea
\eta^{\mu\nu} \langle T_{\mu\nu}(z) V_{\alpha}(x) V'_{\beta}(y) \rangle = \frac{ \delta^2 \langle \mathcal A(z) \rangle}{\delta J^{\alpha}(x) \delta J'^{\beta}(y)} + \langle T^{\mu}_{\mu}(z) V_{\alpha}(x) V'_{\beta}(y) \rangle. 
\label{traceid1}
\eea
The first term on the right-hand side of the equation above defines the residue of the anomaly pole that we have already discussed and isolated in the previous sections. The second term, instead, is the correlation function obtained by inserting the trace of the EMT on the two point functions $\langle V_{\alpha}(x) V'_{\beta}(y) \rangle$ (with the inclusion of terms of gauge fixings and ghosts). This would be the only contribution describing the explicit breaking of the conformal symmetry - in the absence 
of an anomalous breaking induced by the radiative corrections -. It is also evident from the structure of Eq. (\ref{anom}) that the complete anomalous effective action takes contributions from vertex functions with two and three gauge bosons on the external lines, 
due to the $SU(2)$ and $SU(3)$ field strengths $(F_2,F_3)$.

In the context of theories with extra dimensions, the correlator obtained by the insertion of the trace $T^{\mu}_{\mu}$ plays a key role in describing the radiative corrections to the tree-level coupling of $\phi$ to matter. We present here the explicit form of this vertex when the radion couples to on-shell external photons. It is defined as 
\bea
(2\pi)^4 \delta^{(4)}(k-p-q)\mathcal D^{AA}_{\alpha\beta}(p,q) = -i \frac{\kappa}{2}\int d^4z d^4x d^4y \langle T^{\mu}_{\mu}(z) \, A_{\alpha}(x) A_{\beta}(y) \rangle_{amp}e^{-ikz +ipx +iqy}
\label{insert}
\eea
(with an amputated correlation function) and can be decomposed in the form
\bea
\mathcal D^{(AA)\a\b}(p,q) = \mathcal D_{F}^{(AA)\a\b}(p,q) + \mathcal D_{B}^{(AA)\a\b}(p,q) +\mathcal D_{I}^{(AA)\a\b}(p,q),
\label{coup}
\eea
where
\bea
\mathcal D_{F}^{(AA)\a\b}(p,q) &=& - i \frac{\kappa}{2}\frac{\alpha}{\pi} \sum_f Q_f^2 m_f^2 \left[ \frac{4}{s} + 2 \left(\frac{4 m_f^2}{s}-1\right) \mathcal C_0\left(s,0,0,m_f^2,m_f^2,m_f^2 \right)\right]\, u^{\alpha\beta}(p,q) \,, \\
\mathcal D_{B}^{(AA)\a\b}(p,q) &=& - i \frac{\kappa}{2} \frac{\alpha}{\pi} \left[ 6 M_W^2 \left(1-2\frac{M_W^2}{s}\right)  \mathcal C_0\left(s,0,0,M_W^2,M_W^2,M_W^2 \right) - 6 \frac{M_W^2}{s} -1 \right] \, u^{\alpha\beta}(p,q) \,, \label{GBdilaton} \\
\mathcal D_{I}^{(AA)\a\b}(p,q) &=& - i \frac{\kappa}{2} \frac{\alpha}{\pi} \left[ 1 + 2 M_W^2 \mathcal C_0\left(s,0,0,M_W^2,M_W^2,M_W^2 \right) \right] \, u^{\alpha\beta}(p,q) \nn \\
&&+ i\frac{\kappa}{2} \frac{\alpha}{\pi} M_W^2 \frac{s}{2} \mathcal C_0 \left(s,0,0,M_W^2,M_W^2,M_W^2 \right) \, \eta^{\alpha \beta} \, \label{GIdilaton}
\eea
correspond to the contributions coming from the insertion on the photon 2-point function of the trace of the EMT, as specified 
in (\ref{insert}). These correspond 
to fermion ($F$) and boson ($B$) loops, together with terms of improvement ($I$). A description of these terms can be found in \cite{Coriano:2011ti}.\\
Note that these expressions are ultraviolet finite and do not need any renormalization counterterm. One can also observe the presence of two scaleless terms in Eq. (\ref{GBdilaton}) and Eq. (\ref{GIdilaton}), (the $\pm i\frac{\kappa}{2}\frac{\alpha}{\pi s}$ terms),  which do not depend on any mass parameter but only on $1/s$. These are not part of the anomaly - since the 
${\mathcal D}$'s correspond to explicit breaking of the conformal symmetry -  and seem to invalidate 
our argument about the pole origin of the entire anomaly for a spontaneously broken theory. 
However these extra scaleless contributions, as one can easily check, cancel in (\ref{coup}) {\em if the Higgs scalar is conformally coupled}, since they appear with the opposite sign.\\ 
We can summarize this analysis by saying that $\mathcal D^{AA}_{\alpha\beta}$ is zero for a conformal theory (e.g. QED with massless fermions) and it is expected to be proportional to any mass parameter of the theory otherwise. For instance it is nonzero for QCD and QED when the quarks are massive. 
Indeed one can explicitly check, for example, that in the QCD case 
the corresponding amplitude $\mathcal D^{gg}_{\alpha\beta}$, coming from the insertion of the trace of the EMT on the gluon 2-point function, even if not zero, does not contribute any scaleless  term on the right-hand-side of the anomaly equation (Eq (\ref{traceid}) or (\ref{traceid1})). 
The same is true in the electroweak theory only if the Higgs doublet is conformally coupled to gravity. In our case this is guaranteed - by construction - due to the specific choice of the coefficient in front of the term of improvement. If the improvement had not been included in the EMT, then this would have implied that extra scaleless contributions had to combine with the pole term in Eq. (\ref{Phi1Bpole}) to saturate the anomaly. This is equivalent to saying that the pole term in the correlator, in this specific case, would not be entirely responsible for the generation of the anomaly. Indeed, for a conformally coupled Higgs only the sum of Eq. (\ref{Phi1Bpole}) and (\ref{Phi1Ipole}) encloses the entire contribution to the anomaly, which thus can be entirely attributed to the pole part. 

It is important to observe that if the definition of the coupling of the dilaton $\phi$ to the trace of the EMT is 4-dimensional, then there is no tree level coupling of the same state to the anomaly. This is indeed the content of Eq. (\ref{coup}), which does not include 
any anomalous term of the form $\phi FF$ generated by the classical Lagrangian geometrically reduced on the brane. For this reason, the coupling of the dilaton to the anomaly 
is  obtained only if we make one extra assumption. \\ 
For instance, in our formulation we need to replace the $\phi T_\mu^\mu$ vertex appearing in (\ref{comp}) with the vertex 
$\phi g^{\mu\nu}\langle T_{\mu\nu}\rangle $ at the onset, and then use Eq. (\ref{traceid}). Notice that in this expression the EMT does not need to be 
renormalized. In fact, one  can show explicitly that the renormalization does not affect the trace of the same tensor, being the counterterm vertex in $TJJ$ proportional to a traceless form factor \cite{Giannotti:2008cv, Armillis:2010qk}.  For this reason the operation of trace on 
$T^{\mu\nu}$ (i.e. $g^{\mu\nu} \langle T_{\mu\nu}\rangle$) can be computed by the insertion of the bare EMT in 2-point functions. \\
 In other approaches the same coupling requires a redefinition of the trace of the EMT from 4 to $D$ dimensions. In this second case the renormalization of the trace operator is essential in order to generate the coupling of the dilaton to the complete scale violations (anomaly plus explicit terms) present in the anomaly equation. This is obtained by the replacement in (\ref{comp}) of $\phi T^\mu_\mu$ (in 4 dimensions) with $\phi \langle {T_{r}}^{\mu}_{\mu}\rangle_D $), where ${T_{r}}^{\mu}_{\mu}$ is the trace of the renormalized EMT, computed in $D$ dimensions \cite{Giudice:2000av, Csaki:2000zn}.

We are now going to briefly discuss and compare the structure of the effective scalar interactions obtained from the trace anomaly pole against those coming from the exchange of a  fundamental radion introduced by a generic extra dimensional model. An effective degree of freedom in the form of a dilaton ($\varphi F F$) interacting both with the anomaly and with the (explicit) scale violating terms is induced by the effective action generated  by the anomaly loop. This effective interaction can be carefully identified in the 1-graviton exchange channel not only for a massless theory \cite{Giannotti:2008cv} but also in the presence of explicit scale non-invariant terms.
One can investigate the salient features of these interactions by a direct computation. 

 \subsection{Amplitudes for graviton/radion  exchange in the production of two gauge bosons} 

For example, let's consider the production of two photons by a gravitational source characterized by a certain EMT $T'_{\mu\nu}$. The tree-level amplitudes with the exchange of the first modes of the KK towers, namely a massless graviton and a massless dilaton can be formally written as
\bea
\mathcal M_{grav}^{(0)} &=& -\frac{\kappa^2}{4} \bigg[ T'_{\mu\nu} P^{\mu\nu\rho\sigma}(k^2) V_{\rho\sigma\alpha\beta}(p,q) \bigg] \epsilon^{*\,\alpha}(p)
\epsilon^{*\,\beta}(q) \nn\\
&=& -\frac{\kappa^2}{2} \frac{1}{k^2}\bigg[T'^{\mu\nu}V_{\mu\nu\alpha\beta}(p,q) 
- \frac{1}{n-2}{T'^\mu}_{\mu}{V^\rho}_{\rho\alpha\beta}(p,q)\bigg]\epsilon^{*\,\alpha}(p) \epsilon^{*\,\beta}(q) \,, \\
\mathcal M_{dil}^{(0)} &=& -\frac{\kappa^2}{4} \omega^2 \bigg[ {T'^{\mu}}_{\mu} P(k^2) {V^{\rho}}_{\rho\alpha\beta}(p,q) \bigg] \epsilon^{*\,\alpha}(p)
\epsilon^{*\,\beta}(q)
= -\frac{\kappa^2}{4} \frac{\omega^2}{k^2}\bigg[{T'^{\mu}}_{\mu}{V^{\rho}}_{\rho\alpha\beta}(p,q) \bigg] \epsilon^{*\,\alpha}(p) \epsilon^{*\,\beta}(q) \,, \nn\\
\eea
where the $\epsilon(p), \epsilon(q)$ are the polarization vectors of the two final state photons and $-i \frac{\kappa}{2} V^{\rho\sigma\alpha\beta}(p,q)$ is the graviton - two photons vertex ($M_V = 0$)
\bea
-i \frac{\kappa}{2} V^{\rho\sigma\alpha\beta}(p,q) = - i \frac{\kappa}{2} \bigg\{(k_1 \cdot k_2 + M_V^2) C^{\mu\nu\alpha\beta}
+ D^{\mu\nu\alpha\beta}(k_1,k_2) + \frac{1}{\xi}E^{\mu\nu\alpha\beta}(k_1,k_2) \bigg\}
\eea
with
\bea
&& C_{\mu\nu\rho\sigma} = g_{\mu\rho}\, g_{\nu\sigma}
+g_{\mu\sigma} \, g_{\nu\rho}
-g_{\mu\nu} \, g_{\rho\sigma}\,, \nn
\\
&& D_{\mu\nu\rho\sigma} (k_1, k_2) =
g_{\mu\nu} \, k_{1 \, \sigma}\, k_{2 \, \rho}
- \biggl[g^{\mu\sigma} k_1^{\nu} k_2^{\rho}
  + g_{\mu\rho} \, k_{1 \, \sigma} \, k_{2 \, \nu}
  - g_{\rho\sigma} \, k_{1 \, \mu} \, k_{2 \, \nu}
  + (\mu\leftrightarrow\nu)\biggr], \nn \\
&& E_{\mu\nu\rho\sigma} (k_1, k_2) = g_{\mu\nu} \, (k_{1 \, \rho} \, k_{1 \, \sigma}
+k_{2 \, \rho} \, k_{2 \, \sigma} +k_{1 \, \rho} \, k_{2 \, \sigma})
-\biggl[g_{\nu\sigma} \, k_{1 \, \mu} \, k_{1 \, \rho}
+g_{\nu\rho} \, k_{2 \, \mu} \, k_{2 \, \sigma}
+(\mu\leftrightarrow\nu)\biggr]. \,
\eea
$P^{\mu\nu\rho\sigma}(k^2)$ and $P(k^2)$ are the (massless) graviton and dilaton propagators in the de Donder gauge which are given by, in the framework of dimensional regularization ($n=4-\epsilon$),
\bea
i P^{\mu\nu\rho\sigma}(k^2) = \frac{i}{k^2} \bigg[ \eta^{\mu\rho} \eta^{\nu\sigma} + \eta^{\mu\sigma} \eta^{\nu\rho} - \frac{2}{n-2} \eta^{\mu\nu} \eta^{\sigma\rho}\bigg] \,, \qquad i P(k^2) = \frac{i}{k^2} \,.
\eea
Now we consider the one-loop corrections to these expressions and introduce the notation 
\bea
\Gamma^{(AA)}_{\mu\nu\alpha\beta}(p,q) = -i \frac{\kappa}{2} \bar \Gamma^{(AA)}_{\mu\nu\alpha\beta}(p,q) \,, \qquad \mathcal D^{(AA)}_{\alpha\beta}(p,q) = -i  \frac{\kappa}{2} \bar {\mathcal D}^{(AA)}_{\alpha\beta}(p,q) \,,
\eea
in order to factorize the gravitational coupling constant. We obtain
\bea
\mathcal M_{grav}^{(1)} &=& -\frac{\kappa^2}{2} \frac{1}{k^2}\bigg[T'^{\mu\nu} \, \bar \Gamma^{(AA)}_{\mu\nu\alpha\beta}(p,q) 
- \frac{1}{n-2}{T'^\mu}_{\mu} \, \eta^{\rho\si} \, \bar \Gamma^{(AA)}_{\rho\si\alpha\beta}(p,q)\bigg]\epsilon^{*\,\alpha}(p) \epsilon^{*\,\beta}(q)\nn \\
&=& -\frac{\kappa^2}{2} \frac{1}{k^2}\bigg[T'^{\mu\nu} \, \bar \Gamma^{(AA)}_{\mu\nu\alpha\beta}(p,q) 
- \frac{1}{n-2}{T'^\mu}_{\mu} \, \left(\bar {\mathcal D}^{(AA)}_{\alpha\beta}(p,q) + \mathcal A_{\alpha\beta}(p,q)\right) \bigg]\epsilon^{*\,\alpha}(p) \epsilon^{*\,\beta}(q)\,, 
\label{dil0}
\eea

\bea
\mathcal M_{dil}^{(1)} &=&
 -\frac{\kappa^2}{4} \frac{\omega^2}{k^2}\bigg[{T'^{\mu}}_{\mu}  \, \left( 
 \bar {\mathcal D}^{(AA)}_{\alpha\beta}(p,q) + \mathcal{A}_{\alpha\beta}(p,q)\right)  \bigg] \epsilon^{*\,\alpha}(p) \epsilon^{*\,\beta}(q).  \,
\label{dil}
\eea
The $\mathcal A_{\alpha\beta}(p,q)$ term is the anomaly contribution generated by the pole terms in the $\Gamma^{(AA)}(p,q)$ vertex and it is given by
\bea
\mathcal A_{\alpha\beta} \equiv -2 \frac{\beta_e}{e} u_{\alpha\beta}(p,q).
\eea

Notice that in (\ref{dil}) we have retained both the coupling of the dilaton $\phi$ to the explicit (non-conformal) and anomalous terms generated by the Ward identity of the trace anomaly. A similar scalar interaction appears in the graviton channel (proportional to $T'^\mu_\mu)$, as one can easily infer from the right-hand-side of Eq. (\ref{dil0}). Both interactions are, indeed, of dilaton type, being proportional to the complete 
$({\mathcal D}^{(AA)}_{\alpha\beta} + \mathcal{A}_{\alpha\beta})$ trace of the anomaly loop. We will briefly comment on the origin of this effective dilaton interaction.

For this purpose we recall \cite{Giannotti:2008cv, Armillis:2010qk} that in the case of massless QED the effective interaction induced by the trace anomaly takes the form 
\beq
\mathcal{S}\sim \int d^4 x d^4 y\, R^{(1)} \square^{-1}(x,y) F_{\mu\nu}(y) F^{\mu\nu} (y) 
\label{reig}
\eeq
where $R^{(1)}$ denotes the linearized scalar curvature and $F_{\mu\nu}$ is the abelian field strength. A similar result holds for QCD.  As shown in \cite{Giannotti:2008cv} this expression coincides with the long-known anomaly-induced action derived by Riegert \cite{Riegert:1984kt}, which was derived for a generic gravitational field, after an expansion of its expression around the flat spacetime limit. Notice that in terms of auxiliary degrees of freedom (i.e. two scalar fields $(\varphi, \psi')$) which render the action (\ref{reig}) local \cite{Giannotti:2008cv}, extra couplings of the form $\varphi F F$ are automatically induced by the $1/\square$ term. This interaction is indeed present in the equivalent Lagrangian 
\beq
S_{anom} [g,A;\varphi,\psi'] =  \int\,d^4x\,\sqrt{-g}
\left[ -\psi'\square\,\varphi - \frac{R}{3}\, \psi'  + \frac{c}{2} F_{\alpha\beta}F^{\alpha\beta} \varphi\right]\,,
\label{effact}
\eeq
($c=-\beta(e)/(2 e)$)
where $\varphi$ and $\psi'$ are the auxiliary scalar fields. This action does not account for any correction to the 
trace anomaly equation due to the appearance of mass terms. The presence of an explicit breaking of scale invariance due to the
${\mathcal D}^{(AA)}$ term, however,  can be handled by a modification of the $\varphi FF$ interaction present in (\ref{effact}), i.e. the anomaly term. In the presence of an explicit breaking of scale invariance, the effective dilation $\varphi$ couples to the neutral currents of the final state just like the fundamental dilaton $\phi$, which is the content of the Eq. (\ref{dil0}). One can explicitly check the cancellation of possible "double pole" contributions in the s-channel. These could be induced by the (single) pole of the graviton propagator together with the  anomaly pole coming from the triangle loop (present in $\bar \Gamma^{(AA)}_{\mu\nu\alpha\beta}$) (see Eq. \ref{dil0}). This cancellation holds under the condition that the source EMT $T'^{\mu\nu}$ is conserved, as expected. Indeed this is an additional check of the significance of this effective component of the 1-graviton exchange amplitude generated in the presence of a trace anomaly vertex.

\section{Discussion}
There are some comments which are in order concerning the result of this analysis, which complete those obtained in the QED and QCD cases \cite{Giannotti:2008cv,Armillis:2009pq,Armillis:2010qk}. From all these investigations it seems clear that anomaly mediation can be described, in a perturbative framework, as due to the exchange of effective massless scalar degrees of freedom between gravity and the gauge sector.  The physical interpretation of these singularities is probably easier to grasp by a dispersive analysis, at least in the massless case, as discussed in \cite{Giannotti:2008cv} for QED. In the QED case, in fact, this component is generated (diagrammatically) by a virtual graviton decaying into two on-shell collinear (correlated) fermions, which later decay into two on-shell photons. This interpretation follows from the fact that the spectral density of the fermion loop diagrams ($\rho(s)\sim \delta(s)$), which indeed generates a pole,
is obtained by cutting the $graviton \to \gamma\gamma$ amplitude in the $s=(p+q)^2$ channel, thus setting two intermediate fermion lines in the triangle diagram on-shell. The approach is similar to that of the chiral anomaly \cite{Dolgov:1971ri}. The pole is found only in the massless case.

Obviously, a similar interpretation of the origin of this singularity, which is present over the entire light cone ($s\sim 0$) of the $T VV'$ 3-point function, should also hold in the QCD case (in the massless fermion limit) and should extend also to the gluon loop. In fact, in QCD, beside the contribution of the fermion sector, a separate gauge invariant contribution comes from the gluon sector \cite{Armillis:2010qk}. 

It is also important to remark that the direct computation of \cite{Armillis:2009pq, Armillis:2010qk}, here applied to the entire neutral sector of the Standard Model, is not based on a dispersive analysis and gives the exact expression of the effective action at 1-loop. For this reason it allows to establish the presence of these contributions on a more general ground, both in the massless and in the massive cases, and is in agreement 
with the dispersive analysis (in the massless case).
\subsection{Summary}

We have indeed seen, in combination with a previous study for QCD \cite{Armillis:2010qk}, that an explicit computation of the exact 1-loop effective action (at leading order in the combined gravitational ($\kappa$) and gauge coupling expansion ($g$)) shows two fundamental features: \\
1) In a massless gauge theory the breaking of conformal invariance is characterized by a typical $1/\square$ behaviour.  Notice that this does not exclude the possible appearance of other nonlocal terms in the same effective action, such as those proportional to $\log (\square)$ (see the discussion in \cite{Mazur:2001aa}). These additional terms, at least in the case of the chiral anomaly, are generated by the insertion of the triangle diagram into a graph of higher perturbative order \cite{Anselm:1989gi}. Therefore, in the chiral case, they are not part of the triangle diagram (i.e. of the lowest order contribution to the anomaly). The computations in QED and QCD of the trace anomaly are in line with this result and are in agreement with Riegert's anomaly-induced action 
\cite{Riegert:1984kt} in these two theories. This result of ours appears to be also in agreement with the observations in \cite{Mazur:2001aa} to which we refer for further details. \\
2) In a gauge theory in a spontaneously broken phase, such as the electroweak theory, the radiative and the explicit breaking of conformal invariance are separately identifiable in the ultraviolet limit. This result holds even if conformal invariance is broken by the Higgs vev already at tree-level.

Therefore, the two (distinct) kinematical domains characterized by the dominance of the anomaly (via its massless poles) are described, in our notations, 
by a single invariant, $s$, which denotes the virtuality of the external graviton. In particular, point 1), as we have already mentioned, is related to the $s\sim 0$ behaviour of the anomalous contributions of the $TV V'$ vertex (for a massless theory) 
and point 2) to the $s\to \infty$ limit of the same correlator (in the massive case). \\ 
Hence, in the massive case (point 2)), the $1/s$ contribution appears only 
after an asymptotic expansion at large energy of the anomaly vertex, and should be interpreted as its dominant asymptotic 
component in the trace part. As such, this component is not part of the amplitude in the infrared (i.e. it is not present at $s=0$).  In fact, the computation of the residue of the correlator at
$s =0$ shows that this indeed vanishes when masses are present. In the fermion case, for instance, this result is due to cancellations between the pole and the second and third terms of Eq.~(\ref{taavertex}).\\
There is no doubt, however, that the 
$1/s$ term, present in the expansion of the anomalous diagram at large $s$, is a manifestation of the same "anomaly pole" encountered in the infrared in the massless case, since its contribution is asymptotically corrected by mass effects ($M^2/s^2$) which become negligible in the UV limit. It seems obvious that we should recover the behaviour typical of a massless theory as we move to high energy,  and the pole term of the anomaly, as the expansions (\ref{expan}) suggest. We have seen that effective dilaton interactions are automatically part of the effective action which parallel those introduced within models with large extra dimensions.

Thus, it could be of interest, for instance, to explore the role that such contributions could play in the analysis of perturbations in the early universe, for instance in the context of inflation driven by a vector field \cite{Dimopoulos:2009vu}, where such interactions appear to be necessary. In this case this vertex would be a direct consequence of the conformal anomaly, without any need to resort to more complex scenarios for its generation. We have also rigorously shown that a pole term completely accounts for the anomaly, in the Standard Model, only if the Higgs scalar is conformally coupled. In the non-conformally coupled case, extra scaleless contributions appear in the anomalous Ward identity for the $TVV'$. For a conformally coupled scalar indeed there is a cancellation between scaleless contributions coming from the explicit breaking of the conformal symmetry and those generated by 
the terms of improvement (Eqs. (\ref{GBdilaton}) and (\ref{GIdilaton})). Obviously, this picture is typical only of theories with a spontaneous breaking of the gauge symmetry. 
For unbroken gauge theories this subtlety disappears and the pole completely accounts for the anomaly, as found in previous analysis of QED and QCD.

\section{Conclusions}
The computation of the effective action describing the interaction of gravity with the Standard Model, related to the trace anomaly, is described by the diagrams that we have analyzed in this work. This approach, even if rather laborious, allows to derive the exact expression of such an action  at leading order, which is the starting point for further phenomenological analysis. As we have shown,  this is characterized by the presence of effective massless degrees of freedom in two kinematical domains. \\ 
One of the main phenomenological applications of these results is in  theories with large extra dimensions. In this context, we have illustrated rather rigorously that the coupling of a radion to the anomaly requires a specific prescription on the definition of the quantum trace of the energy-momentum tensor for these theories.  We have discussed two different (but equivalent) ways to obtain this interaction using extra dimensional models. We have also shown that the appearance of an effective dilaton - coupled both to the anomaly and to extra scale-dependent terms - is a generic feature of the effective action which accounts for the trace anomaly. 
This effective interaction can be identified in the 1-graviton exchange channel when we couple Einstein gravity in 4 dimensions to the the Standard Model. We have illustrated the analogies between the interaction of the radion and of the effective dilaton using some examples. 

 \centerline{\bf Acknowledgments} 
We thank E. Mottola, R. Armillis, E. Dimastrogiovanni, N. Irges and M. Karciauskas for discussions.


\end{document}